\documentclass[preprintnumbers,amsmath,amssymb,usenatbib,nofootinbib,superscriptaddress,showkeys,showpacs,10pt]{revtex4-2}

\usepackage{amsmath}
\usepackage{amsfonts,color}
\usepackage{amssymb,float}
\usepackage{graphicx}
\usepackage{appendix}
\usepackage[colorlinks]{hyperref}
\usepackage{xcolor}
\usepackage{orcidlink}
\usepackage{epsf}
\usepackage{bm}
\usepackage{epstopdf}
\usepackage{natbib}
\usepackage{lipsum}
\usepackage{multirow}
\usepackage{nicematrix}
\usepackage{subfigure}

\begin{document}

\title{Thermodynamics of Deformed AdS-Schwarzschild Black Holes Beyond the Bekenstein Paradigm}

\author{Jafar Sadeghi\orcidlink{0000-0003-4549-1766}}
\email{pouriya@ipm.ir}
\affiliation{Department of Physics, Faculty of Basic Sciences, University of Mazandaran\\P. O. Box 47416-95447, Babolsar, Iran}
\affiliation{School of Physics, Damghan University, P. O. Box 3671641167, Damghan, Iran}

\author{Fatemeh Khosravani\orcidlink{0009-0008-6905-5305}}
\email{f.khoravaniamiri01@umail.umz.ac.ir}
\affiliation{Department of Physics, Faculty of Basic Sciences, University of Mazandaran\\P. O. Box 47416-95447, Babolsar, Iran}
\affiliation{School of Physics, Damghan University, P. O. Box 3671641167, Damghan, Iran}

\author{Alireza Amani\orcidlink{0000-0002-1296-614X}}
\email{al.amani@iau.ac.ir}
\affiliation{Department of Physics, Am.C., Islamic Azad University, Amol, Iran.}

\vspace{1.5cm}

\begin{abstract}
This work investigates the thermodynamic behavior of deformed AdS-Schwarzschild black holes by incorporating higher-order corrections within non-interacting spacetime models and extended entropy frameworks. To address the inadequacies of classical statistical mechanics in describing gravitational systems with non-local and long-range interactions, we employ non-extensive entropy formalisms, specifically Tsallis and Barrow entropies, which capture quantum-scale deviations and extended correlations. The resulting thermodynamic analysis reveals significant departures from conventional black hole behavior under strong entropy deformations. Notably, as the degree of non-extensivity decreases, the system asymptotically recovers classical features, indicating an emergent universality across statistical regimes. Furthermore, the Joule-Thomson (JT) expansion is examined to analyze the temperature-pressure response during adiabatic processes. Key thermodynamic quantities, including mass, temperature, heat capacity, Gibbs free energy, enthalpy, internal energy, and the JT coefficient, are computed under the influence of non-extensive entropy corrections. These results provide deeper insight into black hole thermodynamics in quantum-corrected spacetimes and offer new avenues for exploring gravitational systems beyond the traditional Bekenstein-Hawking (BH) framework.
\end{abstract}

\date{\today}



\maketitle
\tableofcontents
\newpage

\section{Introduction}

The study of black holes has long captivated scientists not only from a gravitational standpoint but also from a thermodynamic perspective. Black hole thermodynamics, a compelling intersection of classical thermodynamics, general relativity, quantum mechanics, and statistical mechanics, is now regarded as a foundational area of theoretical physics. Pioneering researchers such as Bekenstein, Hawking, Bardeen, and Carter extended thermodynamic principles to black holes, revealing profound implications for our understanding of the universe. Since the seminal work of Bekenstein and Hawking in 1973, which proposed that black holes possess entropy proportional to the area of their event horizon  \cite{Bardeen1973, Hawking1975, Bekenstein1973}, black holes have been recognized as thermodynamic systems that obey the laws of thermodynamics. It is known that for certain black hole systems, a well-defined thermodynamic relation emerges between entropy and temperature, revealing a distinctive connection that sets them apart from conventional thermodynamic systems. \cite{Khosravani2024}. The thermodynamic behavior and stability of various black hole solutions have been extensively explored in the literature \cite{Mele2022, Munch2023, Xiao2022, Calmet2021, Ganai2019, Panotopoulos2021}. Building on Bekenstein’s insights, S. Hawking introduced the concept of Hawking radiation in 1974, demonstrating that black holes emit radiation with a spectrum analogous to that of a black body at a specific temperature. This temperature is proportional to the surface gravity of the black hole. Hawking’s discovery challenged the classical notion that nothing can escape a black hole, showing instead that black holes lose mass and energy through radiation. These findings led to the formulation of the four laws of black hole thermodynamics, which closely mirror the classical laws of thermodynamics. Consequently, black hole thermodynamics has gained significant traction, incorporating new parameters such as heat capacity, pressure, and chemical potential \cite{Bekenstein1980}.

A central tenet of black hole thermodynamics is the BH area law, which asserts that the entropy of a black hole is proportional to the area of its event horizon. However, this law is primarily valid for large black holes in equilibrium\cite{Bekenstein1973}. For small black holes, where Hawking radiation plays a dominant role, the area law becomes insufficient, as it neglects long-range interaction effects. In strong gravitational fields—such as those near black holes—these interactions cannot be ignored, even if they occur over large distances. As a result, Boltzmann–Gibbs statistics may not be the most appropriate framework for describing entropy in such regimes. This limitation was first noted by Gibbs, who addressed systems with divergent partition functions. Systems influenced by strong gravity fall outside the scope of Boltzmann–Gibbs theory \cite{Chavanis2020}. Moreover, several unresolved issues in black hole physics, including thermodynamic stability, are inherently non-extensive. To address these challenges, non-extensive entropy frameworks have emerged in recent decades, offering quantum corrections derived from various approaches. These corrections have significantly impacted the thermodynamic analysis of rotating and charged black holes in diverse spacetimes, including BTZ black holes, massive AdS black holes, and dilatonic black holes \cite{Upadhyay2018a, Upadhyay2019a, Upadhyay2018b, Pourhassan2018}. The influence of corrected entropy has also been extended to specific solutions such as the Schwarzschild–Beltrami-de Sitter and Goedel black holes \cite{Upadhyay2019b, Islam2019}. For instance, it has been shown in \cite{Amigo2018} that a proper generalization of entropy is essential for consistent thermodynamic expansion. Non-extensive entropies are derived from several parametric extensions of the Boltzmann–Gibbs statistical entropy formula, which are more suitable for describing complex systems with long-range interactions. These entropies form the foundation of extended statistical frameworks, with significant implications for cosmological environments  \cite{Davies1977, Kaburaki1993, Kaburaki1996} and modified gravity theories  \cite{Zubair2021, Aditya2019}. In recent years, significant progress has been made in understanding regular and quantum-corrected black hole geometries, energy conditions, and causal structure within modified spacetime frameworks, where both Lorentzian and Euclidean formulations play an important role in resolving singularities and exploring physical viability \cite{Wang2026a, Wang2026b, Capozziello2024, Battista2024}.

Recent efforts have focused on understanding the impact of the absence of expansion on the phase structure of black holes \cite{Volonteri2021}. The advantages of such an expansion are evident in the small–large charged AdS black hole transition (in the canonical ensemble), which closely resembles the phase transition of a van der Waals fluid  \cite{Takata2021}. Among these cases, this paper investigates the thermodynamics of AdS–Schwarzschild black holes using Tsallis and Barrow non-extensive entropy formalisms. A standard and simplified approach to describing black hole thermodynamics involves electromagnetic fields and a positive or negative cosmological constant, both of which influence thermodynamic quantities such as entropy and temperature \cite{Anand2024, Dabrowski2024look}. The study of microscopic structures and stability of black holes within the AdS/CFT framework has further deepened our understanding of these complex systems. AdS black holes correspond to thermal states in conformal field theory (CFT), and this correspondence, known as holographic thermodynamics, has garnered considerable attention \cite{Ladghami2025, Ageev2025, Ladghami2024, Maldacena1999}. In the context of the Taub–NUT black hole, the left and right central charges are found to be equal. This degeneracy carries significant consequences for the AdS/CFT correspondence, influencing both the holographic duality framework and the thermodynamic characterization of black hole solutions \cite{Sadeghi2023}. Recent developments have also explored the effects of quantum fluctuations on black holes in general relativity,  various modified gravity theories, and white hole \cite{DelaCruz2012, Khlopov1985, Kubeka2022}. These investigations often yield modified expressions for black hole entropy and temperature, which may have observable consequences and offer deeper insights into quantum gravity \cite{Anacleto2015, Artemovych2020}. Thus, black hole thermodynamics remains a vital tool for probing the mysteries of quantum gravity and the fundamental nature of space-time \cite{Yang2024, Ashtekar2004, Ali2009}. 

Recently, topology-based thermodynamics has gained significant attention in the study of black hole phase transitions and critical phenomena. Rooted in topological current theory, this framework captures global and invariant properties of the thermodynamic phase space beyond local fluctuations. Phase transitions are interpreted as topological transformations within the underlying geometric structure, with critical points linked to changes in topological charge. The approach identifies stable and unstable configurations through the curvature and continuity of the thermodynamic potential, while variations in the topological structure of the free energy landscape distinguish first- and second-order transitions. By unifying geometric and thermodynamic principles, it offers deeper insight into the stability and evolution of black hole systems. Overall, the topological perspective reveals fundamental connections between geometry, stability, and phase structure in gravitational thermodynamics \cite{Wei2022, Sadeghi2023Bardeen, Sadeghi2024Bulk, Wu2024Multi, Sadeghi2024Hayward, Sadeghi2024Hyperscaling, Wu2023TaubNUT, Hazarika2024, Gashti2024, Afshar2024Role, Afshar2024Potential, Panah2024, Wu2025NUT, Sadeghi2024Kiselev, Brzo2025, Afshar2025CFT, Gashti2025PhaseSpace, Alipour2023, Anand2025, Gashti2025Holographic, Afshar2025WGC, Heidari2025, Wei2020, Cunha2017, Afshar2025Aschenbach, Afshar2025PhotonSphere, Alipour2025WGC}. In addition, detailed studies of geodesic motion, null geodesics, black hole shadows, and dynamical accretion processes in such effective geometries have provided important insights into observational signatures and the thermodynamic interpretation of quantum-corrected black holes \cite{Wang2025, Capozziello2025, Amani2012a, Amani2012b, Battista2026Shadow, Battista2022Geodesic}.

Within the framework of extended phase space thermodynamics, the equation of state and thermodynamic stability of black holes have become indispensable tools for understanding gravitational phase transitions. In this framework, the cosmological constant is interpreted as a thermodynamic pressure, and its conjugate volume plays the role of the thermodynamic volume of the black hole. This interpretation establishes a profound analogy between AdS black holes and ordinary thermodynamic systems, particularly van der Waals fluids, leading to the discovery of rich critical phenomena, first- and second-order phase transitions, and critical points \cite{Kubiznak2012, Dolan2011, Altamirano2014}. Furthermore, the analysis of thermodynamic stability through heat capacity, isothermal compressibility, and other thermodynamic responses provides valuable information about the stable and unstable regions of the parameter space \cite{Davies1989, Chamblin1999a, Chamblin1999b}. In recent years, it has been shown that modifications due to generalized entropies can significantly change the structure of the equation of state, the stability boundaries, and the critical behavior of black holes \cite{Majhi2017, Pradhan2019}. Therefore, investigating the equation of state together with different stability criteria within the Tsallis and Barrow entropy formalisms provides a natural framework for understanding how statistical and geometrical corrections influence the phase structure, stability, and thermodynamic behavior of deformed AdS–Schwarzschild black holes.

Motivated by these developments, it has become evident that the conventional Hawking entropy is inadequate for capturing the full thermodynamic behavior of gravitational systems. Consequently, considerable attention has been devoted to generalized entropy formalisms that extend the standard Hawking prescription \cite{Sadeghi2025PLEB, Sadeghi2026NPB}. To the best of our knowledge, a systematic investigation of the equation of state, thermodynamic stability, and JT expansion of deformed AdS–Schwarzschild black holes within both Tsallis and Barrow non-extensive entropy frameworks has not yet been reported. The motivation behind this work is to examine the influence of thermal fluctuations on the thermodynamic properties of a deformed AdS black hole system \cite{Gogoi2025, Bardeen1973}, and to analyze the stability and behavior of these systems under such perturbations. We begin by constructing the fundamental thermodynamic variables of the AdS black hole system. Then, employing the non-extensive Tsallis and Barrow entropy frameworks, we compute and analyze the mass, temperature, specific heat, and Gibbs free energy, each of which reveals subtle features of thermal equilibrium. The JT effect, which describes how the temperature of a gas changes during a pressure variation process, identifies the critical point of transition between temperature increase and decrease. This behavior is quantified by the JT coefficient. Previous studies have examined this process for various AdS black hole configurations and across different gravitational theories \cite{Mo2020, Mo2018, Bi2021, Ahmed2018}.
 
In this work, we investigate several thermodynamic properties of the deformed AdS–Schwarzschild black hole, with particular emphasis on the JT effect, by employing Tsallis and Barrow non-extensive entropy formalisms. Our analysis shows that, for deformed AdS–Schwarzschild black holes, both Tsallis and Barrow entropies provide a more accurate and physically meaningful description of the system compared to the traditional Boltzmann entropy framework. To illustrate these differences, we present comparative plots for each entropy model, highlighting deviations from the standard Boltzmann case.

The introduction of deformation into the AdS black hole geometry significantly modifies its phase transition behavior and, consequently, its thermal stability. These modifications demonstrate the sensitivity of black hole thermodynamics to non-classical effects and underscore the importance of non-extensive entropy in capturing such phenomena. Such investigations not only enhance our understanding of the thermodynamic behavior of deformed black holes but also provide insights into the interplay between horizon geometry, entropy generalizations, and stability in AdS spacetimes.

The paper is structured as follows. In Section \ref{II}, the thermodynamic framework of the deformed AdS–Schwarzschild black hole is introduced and the main thermodynamic quantities are derived in the framework of the generalized Tsallis and Barrow entropies. Then, the effect of these two approaches on heat capacity, Gibbs free energy, enthalpy, internal energy, phase transitions, and the JT process is investigated. In Section \ref{III}, the black hole equation of state is extracted and the thermal stability, mechanical stability, and behavior of the system in the extended phase space are analyzed. Finally, in Section \ref{IV}, the main results of the research are summarized and, by comparing the Tsallis and Barrow entropy frameworks, the physical implications of the findings for the thermodynamics of deformed black holes, phase transitions, thermodynamic stability, and future research directions in the modified gravity framework are discussed.


\section{Deformed AdS-Schwarzschild black hole}\label{II}
In \cite{Khosravipoor2023}, it was demonstrated that by applying the gravitational decoupling method, one can obtain a deformed AdS–Schwarzschild black hole solution in the presence of an additional gravitational source that satisfies the weak energy condition. In this construction, the energy density of the additional source is chosen deliberately as a monotonic function, ensuring consistency with the required physical constraints. The method introduces a positive deformation parameter, which controls the strength of the geometric deformations imparted to the background spacetime. Importantly, the requirement that the solution possesses an event horizon imposes an upper bound on the value of this deformation parameter. After deriving the relevant thermodynamic quantities as functions of the event horizon radius, the analysis primarily focuses on how variations in the deformation parameter influence the horizon structure, the overall thermodynamics of the black hole, and the temperature of the Hawking–Page phase transition. The study reveals several key effects of increasing the deformation parameter: the minimum horizon radius required for a black hole to achieve local thermodynamic equilibrium decreases, the minimum temperature below which no black hole exists is reduced, and both the horizon radius and the temperature associated with the first-order Hawking–Page phase transition increase. Moreover, in the limit where the deformation parameter vanishes, the thermodynamic behavior of the black hole reduces smoothly to the standard results reported in the literature for undeformed AdS–Schwarzschild black holes, confirming the consistency of this approach. These findings highlight the significant role of geometric deformations in modifying black hole thermodynamics and phase transition characteristics, providing a deeper understanding of the interplay between spacetime geometry and black hole thermal properties.
According to \cite{Khosravipoor2023}, the warped AdS black hole metric corresponds to:\\
\begin{equation}\label{1}
ds^{2}=-f(r)dt^{2}+f^{-1}(r)dr^{2}+r^{2}d\Omega^{2},
\end{equation}
By substituting the energy density \cite{Gogoi2025} into the field equations, the metric function of this black hole is as follows:
\begin{equation}
f(r)=\frac{r^2}{\ell^2}-\frac{2 M}{r}+\frac{\alpha  \left(\beta ^2+3 r^2+3 \beta  r\right)}{3 r (\beta +r)^3}+1,
\end{equation}
In which $\beta$ is a constant parameter that controls the behavior of energy density at zero radius, and $\alpha$ is the deformation parameter. Here, $\ell$ is the AdS radius in terms of the cosmological constant as follows.
$
\ell^2=-\frac{3}{\Lambda }.
$
This metric function describes a black hole solution with a deformation parameter $\alpha$, which preserves the asymptotic AdS behavior and introduces corrections proportional to additional gravitational fields. The horizon structure and thermodynamic properties of this deformed AdS black hole may not correspond to a regular black hole or may lack a safeguarded singularity. It can be clearly observed that the expressions related to the black hole mass $M$ and the deformation parameter exhibit singularity issues, where the $M$ determines the black hole mass and $r_h$ is the horizon radius. The black hole mass can be obtained by, $f(r)=0$.

The mass, the temperature, and the conjugate volume in terms of horizon radius $r_h$ in the deformed AdS-Schwarzschild black holes are given as
\begin{eqnarray}
&M=\frac{1}{2} \left(-\frac{\Lambda  r_h^3}{3}+\frac{\alpha  \left(\beta ^2+3 r_h^2+3 \beta  r_h\right)}{3 (\beta +r_h)^3}+r_h\right),\label{M11}\\
&T= -\frac{\alpha r_h}{4 \pi  \left(\beta+r_h\right)^4}+\frac{1}{4 \pi r_h} -\frac{\Lambda r_h}{4 \pi},\label{T11}\\
&V=\frac{4 \pi  r_h^3}{3}.\label{V1}
\end{eqnarray}

As explained in the introduction, extensive entropy, which is BH entropy, while playing an important role in thermodynamics, ignores long-range forces. Some physical systems, because they are affected by such forces, cannot be properly described by Gibbs thermodynamics. Now, we examine the role of two of these long-range entropies, named Tsallis and Barrow, on deformed AdS-Schwarzschild black hole to see what effect they have on the thermodynamics of these black holes.
\subsection{Tsallis entropy}
 There are certain physical systems for which Gibbs thermodynamics may not be an appropriate choice, because they are influenced by long-range forces such as black holes. For this reason, here we consider the Tsallis entropy for non-extensive systems to address this issue. The form of Tsallis entropy is as follows \cite{Ghaffari2024}:
\begin{equation}\label{7}
S_{T}=\gamma (S_{BH})^{\delta}=\gamma (\pi r_{h}^{2})^{\delta},
\end{equation}
where $\delta$ is the entropy parameter which is positive, and $\gamma$ is the normalization coefficient, both of these parameters quantify the deviation from the BH entropy.

The mass of this black hole, in terms of the Tsallis entropy from relations \eqref{M11} and \eqref{7}, is given by:
\begin{equation}\label{8}
M = \frac{\alpha}{6 a^3} \left(\beta ^2+3 \beta  \left(a-\beta\right) + 3 \left(a-\beta\right)^{2}\right) - \frac{1}{6} \Lambda \left(a-\beta\right)^{3} + \frac{1}{2} \left(a-\beta\right),
\end{equation}
where $a=\beta +\left(\frac{\pi^{-\delta} S_T}{\gamma }\right)^{\frac{1}{2\delta}}$. We can also obtain the temperature of this black hole in terms of the Tsallis entropy using relations \eqref{T11} and \eqref{7} as follows:
\begin{equation}\label{Ttes1}
T = -\frac{a-\beta}{4 \pi} \left(\Lambda+\frac{\alpha}{a^4}\right) + \frac{1}{4 \pi (a-\beta)},
\end{equation}
where the pressure parameter is considered constant and is proportional to the cosmological constant as
\begin{equation}\label{11}
\Lambda =-8 \pi  p.
\end{equation}

We now proceed to calculate the heat capacity, so that its positive sign is sufficient to ensure stability. Therefore, for $C_{P}>0$, the black hole system is stable, whereas $C_{P}<0$ indicates instability. In that case, the heat capacity is as
\begin{equation}\label{10}
C_{P}=T \left(\frac{\partial S_T}{\partial T}\right)=T \left(\frac{\partial S_T}{\partial r_{h}}\frac{\partial r_{h}}{\partial T }\right),
\end{equation}
where, by substituting Eq. \eqref{11} into \eqref{Ttes1} and then into Eq. \eqref{10}, the heat capacity at constant pressure in terms of Tsallis entropy is obtained as follows:
\begin{eqnarray}
C_{P} = -2 \pi  \frac{\beta ^4 + 4 \beta ^3 \left(a-\beta\right) + \left(8 \pi  P a^4-\alpha+6 \beta^2\right) \left(a-\beta\right)^{2} + 4 \beta  \left(a-\beta\right)^{3}+\left(a-\beta\right)^{4}}{-8 \pi  P a^4 + \frac{\alpha \left(4 \beta -3 a\right)}{a} + \frac{a}{\left(a-\beta\right)^{2}}}.
\end{eqnarray}

The study of the heat capacity provides a clear insight into the thermodynamic behavior of deformed AdS-Schwarzschild black holes in the framework of the Tsallis entropy. The analysis of the behavior of \( C_p \) not only allows the identification of the regimes in which the phase transition occurs, but also confirms the consistency between the thermodynamic stability and the results obtained from the topological approaches. This agreement between the thermal and topological properties highlights the validity and reliability of the thermodynamic framework when applied to modified gravity models.

\begin{figure}[t]
\begin{center}
\includegraphics[scale=0.8]{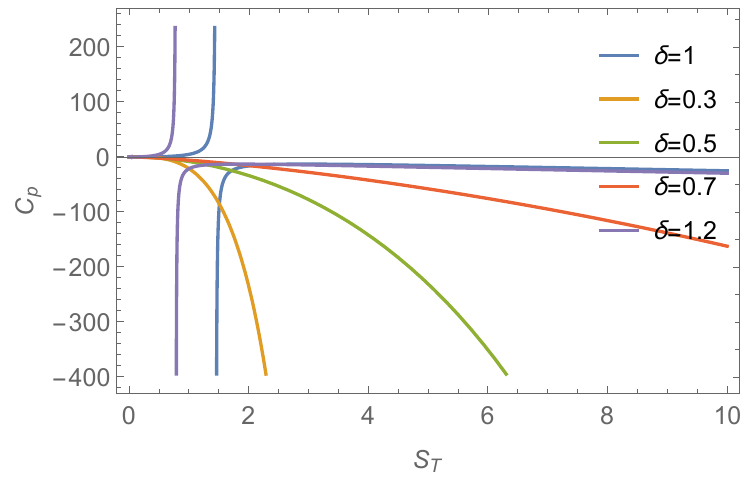}
\caption{Heat capacity $C_p$ of the deformed AdS–Schwarzschild black hole as a function of the Tsallis entropy $S_T$ for several deformation parameters $\delta$ with the fixed model parameters $\alpha=0.45$ and $\beta=0.7$.}
\label{CpT}
\end{center}
\end{figure}

The variations of the heat capacity \( C_p \) in terms of the Tsallis entropy for different values of the black hole parameters show distinct behaviors, as can be seen in Fig. \ref{CpT}. These behaviors can be summarized in two main aspects:

\textbf{Phase transitions:} The divergence of the heat capacity at certain critical values of the horizon radius or control parameters indicates a second-order phase transition. These points represent structural changes in the thermodynamic phase space of the black hole and mark the boundary between stable and unstable regimes.

\textbf{Thermal stability:} Regions with positive heat capacity \((C_p > 0)\) correspond to stable configurations, that is, the system returns to equilibrium after small perturbations. In contrast, negative heat capacity \((C_p < 0)\) indicates thermal instability, which is a common feature of many classical black hole solutions.

In Fig. 1, the behavior of the heat capacity \( C_p \) in terms of the entropy \( S_T \) is investigated for different values of the deformation parameter \( \delta \). The free parameters of the model are set to \( \alpha = 0.45 \) and \( \beta = 0.7 \). The results show that for smaller values of \( \delta \) (such as 0.3 and 0.5), the heat capacity diverges in a region of \( S_T \) and then tends to negative values, this behavior indicates the occurrence of a phase transition and a change in the stability of the black hole. In contrast, for larger values of \( \delta \) (such as 0.7 and 1.2), the curves behave smoother and more stable and the divergences move towards larger values of entropy. This shift indicates that increasing the deformation parameter expands the thermodynamic stability region and makes the black hole more resistant to thermal fluctuations. Overall, this analysis shows that the deformation parameter \( \delta \) plays a crucial role in controlling the stability and phase transitions of the black hole. Also, the Tsallis entropy framework is able to provide a intensive and modified behavior of the heat capacity compared to the Boltzmann–Gibbs entropy and reveals a richer thermodynamic structure.

In what follows, we calculate the Gibbs free energy ($G = M - T S_T$) in terms of Tsallis entropy according to Eqs. \eqref{7}, \eqref{8}, and \eqref{Ttes1} in the following form
\begin{eqnarray}\label{GT1}
&G = \frac{\alpha}{6 a^3}  \left(\beta ^2+3 a  \left(a-\beta\right)\right) + \frac{1}{6} \left(a-\beta\right) \left(3 - \Lambda  \left(a-\beta\right)^{2}\right) + \frac{S_T}{4 \pi } \left(\left(a-\beta\right) \left(\Lambda + \frac{\alpha}{a^4}\right) - \frac{1}{\left(a-\beta\right)}\right),
\end{eqnarray}
where is a function of the Tsallis entropy and the free parameters. Integrating the first law yields the expression for the enthalpy, so, we can calculate the enthalpy in terms of Tsallis entropy in the form
\begin{equation}\label{HT2}
H=\frac{1}{6} \left( \left(-\beta ^3 \Lambda + \frac{\alpha  \beta^2 }{a^3} + 3 \beta\right) + \left(a-\beta\right) \left(\frac{3 \alpha }{a^2}+3\right)-\Lambda  \left(a-\beta\right)^3\right),
\end{equation}
where is in terms of Tsallis entropy and free parameters. The internal energy for this black hole in terms of Tsallis entropy according to $U=H-PV$ is given by:
\begin{equation}\label{UT1}
U = \frac{\alpha}{6 a^3}  \left(\beta ^2+3 \beta  \left(a-\beta\right) + 3 \left(a-\beta\right)^{2}\right) + \frac{a}{2},
\end{equation}
where this also becomes a function of the Tsallian entropy.

Here, we present a detailed investigation of the JT expansion as applied to deformed AdS-Schwarzschild black holes with generalized entropy. The JT expansion, a classical thermodynamic process, provides valuable insight into the temperature response of a system undergoing adiabatic expansion. Analogous to conventional thermodynamics, the JT process involves the throttling of a non-ideal fluid, such as a gas or liquid, through a valve or porous plug, transitioning from a region of higher pressure to one of lower pressure. Importantly, this expansion occurs without heat exchange with the surroundings, making it an adiabatic process. Despite its inherent irreversibility, the enthalpy of the system remains constant, characterizing it as an isenthalpic process. This property allows the study of temperature variations that arise solely from changes in pressure and volume under constant enthalpy conditions.

In the context of black hole thermodynamics in extended phase space, the enthalpy is naturally identified with the black hole mass, enabling a close analogy with conventional thermodynamic systems where enthalpy governs processes involving both work and energy exchange. The JT coefficient quantifies the temperature response to pressure changes during isenthalpic expansion or compression. A positive JT coefficient indicates that the black hole cools upon expansion, whereas a negative coefficient corresponds to heating. The transition between cooling and heating regimes is defined by the inversion curve, which separates these behaviors in the temperature–pressure plane. Identifying these regimes provides insight into the interplay between gravitational dynamics and thermodynamic stability and reveals whether energy is absorbed or released during expansion.

This analysis demonstrates that the JT effect serves as a diagnostic tool in extended black hole thermodynamics. It reveals the thermodynamic behavior of black holes under isenthalpic processes and establishes a conceptual link between classical fluid systems and gravitational thermodynamics, highlighting connections between microscopic physics, horizon geometry, and macroscopic stability. Applying this formalism to deformed AdS-Schwarzschild black holes with generalized entropy allows for a deeper understanding of their thermodynamic behavior under the JT expansion. So, we will have:
\begin{equation}\label{12}
\mu=\left(\frac{\partial T}{\partial P}\right)_{M},
\end{equation}
where the temperature variation during JT expansion is characterized by the JT coefficient and is denoted by $\mu$, which represents the rate of temperature variation with respect to pressure while keeping the enthalpy constant. Using the thermodynamic relations of the black hole and employing Maxwell's relation and heat capacity at constant pressure, this coefficient can be rewritten in the following equivalent form \cite{Okcu2017}
\begin{equation}\label{12-1}
\mu=\frac{1}{C_{P}}\left[T \left(\frac{\partial V}{\partial T}\right)_{P,Q}-V\right],
\end{equation}
where this relation serves as the basis for JT analyses in modified gravity models and generalized entropy frameworks.

Therefore, for our model, the influence of Tsallis entropy will be calculated using the following expression:
\begin{eqnarray}\label{mubar1}
&\mu = \frac{4 \left(a -\beta \right) \left(\Lambda  \,a^{7} - 2 \Lambda  \beta \,a^{6} + \left(\Lambda  \,\beta^{2}-2\right) a^{5} + 3 \alpha a^{3} - 8 \alpha  \beta a^{2} + 7 \alpha  \,\beta^{2} a - 2 \alpha  \,\beta^{3}\right)}{3 a \left(\Lambda  \,a^{6}-2 \Lambda  \,a^{5} \beta +\left(\Lambda  \,\beta^{2}-1\right) a^{4}+ \alpha a^{2} -2 \alpha  \beta a +\alpha  \,\beta^{2}\right)},
\end{eqnarray}
where it depends on the Tsallis entropy and the free parameters.

\begin{figure}[t]
\begin{center}
\includegraphics[scale=0.8]{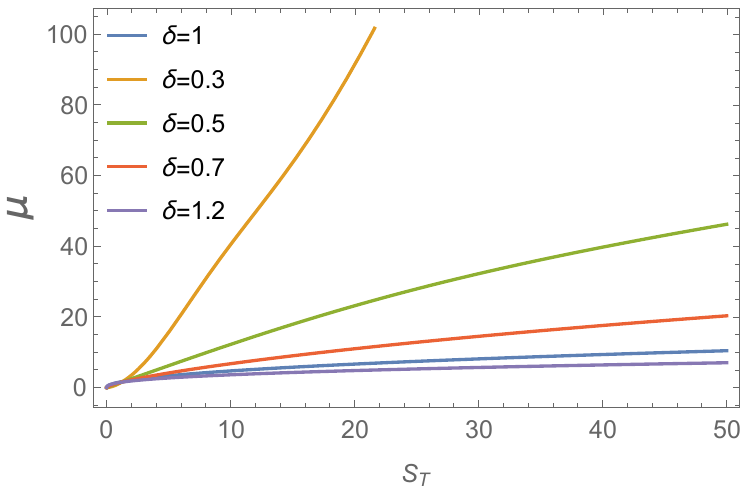}
\caption{The JT expansion coefficient $\mu$ of the deformed AdS–Schwarzschild black hole in terms of the Tsallis entropy $S_T$ for several deformation parameters $\delta$ with the fixed model parameters $\alpha=0.45$ and $\beta=0.7$. }
\label{MuT}
\end{center}
\end{figure}

In Fig. \ref{MuT}, the behavior of the JT expansion coefficient \(\mu\) for a deformed AdS-Schwarzschild black hole in the framework of the Tsallis entropy is investigated in terms of the intensive entropy \(S_T\) and for different values of the deformation parameter \(\delta\). The horizontal axis represents \(S_T\) and the vertical axis represents \(\mu\), and each curve corresponds to a certain value of \(\delta\). The Fig. \ref{MuT} actually reveals the role of the generalized entropy and the deformation parameter in the JT process of the black hole. From the perspective of black hole thermodynamics, the JT coefficient is the main indicator of the cooling or heating behavior of the black hole under adiabatic expansion at constant enthalpy. The regions with \(\mu>0\) correspond to the cooling regime (decreasing temperature with decreasing pressure) and the regions with \(\mu<0\) correspond to the heating regime. In the present figure, all the curves in the range indicated by \(S_T\) are in the positive \(\mu\) region and with increasing entropy, the value of \(\mu\) also increases, in particular, for \(\delta=0.3\) the growth of \(\mu\) is much faster than for other values of \(\delta\). This behavior shows that in the framework of the Tsallis entropy, the black hole is in the JT cooling regime over a wide range of entropy and the intensity of this cooling depends sensitively on the deformation parameter. Comparing the curves for different values of \(\delta\) shows that decreasing \(\delta\) (e.g. \(\delta=0.3\)) leads to a significant increase in the JT coefficient at larger entropies, in other words, for smaller deformations, the temperature response of the black hole is stronger than the pressure change at constant enthalpy. In contrast, for larger values of \(\delta\) (e.g. \(\delta=0.7\) and \(\delta=1.2\)) the growth of \(\mu\) is gentler and more controlled, and the curves behave more smoothly and regularly. This qualitative difference suggests that the deformation parameter not only affects the thermodynamic stability (via the heat capacity), but also regulates the intensity and structure of the JT process, in such a way that \(\delta\) can be used as a control parameter for engineering cooling and heating regimes in the black hole parameter space.

From a physical point of view, the increase of \(\mu\) with \(S_T\) can be interpreted as meaning that in the Tsallis entropy framework, the black hole becomes more sensitive to pressure changes in states with higher entropy (which corresponds to a larger thermodynamic volume or a larger horizon radius), and that decreasing pressure at constant enthalpy leads to a more effective temperature decrease. This behavior is consistent with the intensive structure of the Tsallis entropy and suggests that the statistical corrections due to this entropy lead to a richer Joule–Thomson behavior than in the Boltzmann–Gibbs case.

In general, the Fig. \ref{MuT} shows that: a) The Joule–Thomson coefficient in the Tsallis entropy framework for a AdS-Schwarzschild black hole is positive and increasing over a wide range of entropy. b) The deformation parameter \(\delta\) plays a decisive role in the intensity of the Joule–Thomson cooling, such that smaller values of \(\delta\) lead to a stronger temperature response and larger values of \(\delta\) lead to a milder and more stable behavior. c) This regular and traceable dependence \(\mu(S_T,\delta)\) indicates the internal consistency of the thermodynamic model and its ability to describe non-trivial processes in modified gravity and generalized entropy.

\subsection{Barrow entropy}

As a modification of the classical entropy of a black hole, the Barrow entropy arises from the effects of quantum gravity, which deform the event horizon and create a geometry with fractal properties. These quantum distortions change the thermodynamic description of the black hole and lead to the definition of a generalized relation of the BH entropy. This relation is expressed as follows:
\begin{equation}\label{SBA}
S_{B} = \left( \frac{A}{A_{P}} \right)^{\frac{1+\Delta}{2}},
\end{equation}
where \(A\) is the area of the event horizon, \(A_{P}\) is the Planck area, and \(\Delta\) is a dimensionless deformation parameter that quantifies the intensity of the effects of quantum gravity. The range of variation of this parameter is in the range \(0 \leq \Delta \leq 1\); for \(\Delta = 0\), the relation reduces to the standard BH entropy, which corresponds to a smooth horizon with no fractal modifications, while \(\Delta = 1\) represents the most deformation and complex fractal structure of the event horizon.

By substituting the horizon area \(A = 4\pi r_h^2\), the Barrow entropy is rewritten as
\begin{equation}\label{18}
S_B = \left(\pi r_h^2\right)^{\frac{\Delta }{2}+1}.
\end{equation}

Non-expansive entropy frameworks, such as the Barrow entropy, are a valuable extension of classical thermodynamics, allowing for a more detailed study of black hole systems in which quantum and gravitational effects dominate. These approaches provide new theoretical insights into the statistical and microscopic nature of spacetime and have been widely used in recent studies \cite{Ghaffari2024}.

The mass of a black hole in the Barrow entropy framework is given by:
\begin{equation}\label{19}
M = \frac{1}{6 \pi ^{3/2}}\left[\frac{\pi ^2 \alpha }{b^3} \left(\pi \beta ^2 + 3 b \left(b-\sqrt{\pi} \beta\right)\right) - \Lambda \left(b-\sqrt{\pi} \beta\right)^3 + 3 \pi\left(b-\sqrt{\pi} \beta\right) \right],
\end{equation}
where \(b=\sqrt{\pi } \beta +S_B^{\frac{1}{\Delta +2}}\). The temperature and the heat capacity of the black hole in terms of the Barrow entropy are obtained as follows:
\begin{eqnarray}
&T=\frac{1}{4 \pi ^{3/2}}\left[-S_B^{\frac{1}{\Delta +2}} \left(\frac{\pi ^2 \alpha }{b^4}+\Lambda \right)+\pi  S_B^{-\frac{1}{\Delta +2}}\right],\label{TBar1}\\
&C_{P} = -2 \frac{\pi ^2 \beta ^4 + 4 \pi ^{3/2} \beta ^3 S_B^{\frac{1}{\Delta +2}} + \left(8 P b^4 - \pi  \alpha + 6 \pi  \beta ^2 \right) S_B^{\frac{2}{\Delta +2}} + 4 \sqrt{\pi } \beta  S_B^{\frac{3}{\Delta +2}}+S_B^{\frac{4}{\Delta +2}}}{\frac{\alpha \pi}{b} \left(4 \sqrt{\pi} \beta - 3 b\right) + b^4  S_B^{-\frac{2}{\Delta +2}} - 8 b^4  P},\label{21}
\end{eqnarray} 
where the dependence on the Barrow entropy and free parameters is evident.

\begin{figure}[t]
\begin{center}
\includegraphics[scale=0.8]{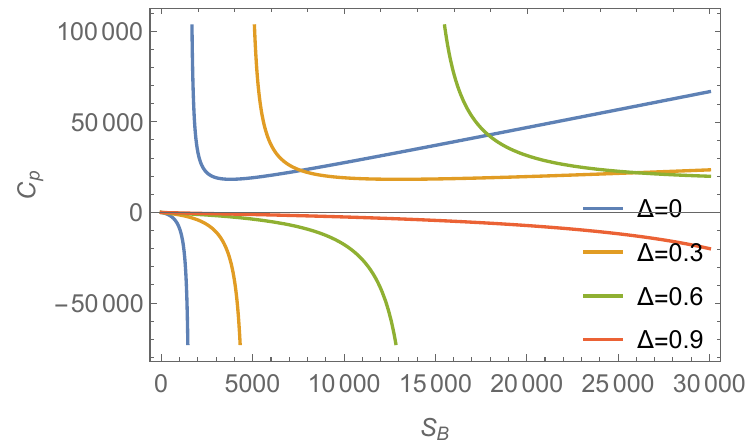}
\caption{Heat capacity $C_p$ of the deformed AdS–Schwarzschild black hole as a function of the Barrow entropy $S_B$ for several deformation parameters $\delta$ with the fixed model parameters $\alpha=0.45$ and $\beta=0.7$. }
\label{CpB}
\end{center}
\end{figure}

In the Fig. \ref{CpB}, the behavior of the heat capacity \(C_P\) of a deformed AdS-Schwarzschild black hole in the Barrow entropy framework is plotted in terms of the Barrow entropy \(S_B\) and for several different values of the deformation parameter \(\Delta\). The horizontal axis shows the Barrow entropy \(S_B\) over a wide range of values from near zero to about \(30000\), and the vertical axis presents the heat capacity at constant pressure. Each curve corresponds to a specific value of \(\Delta\) and its variations reflect the stability structure and phase transitions of the black hole in the presence of the fractal geometry of the event horizon. In the region of small entropies, the curves usually show a more unstable behavior; The heat capacity can be negative or tend towards divergence. These regions represent regimes in which the black hole is thermodynamically unstable and small changes in the environmental conditions can lead to a phase transition or a qualitative change in the equilibrium state. As the entropy increases, the heat capacity gradually approaches positive and stable values, and the curves become smoother and more regular. This transition from the unstable region (with \(C_P<0\) or divergence) to the stable region (with \(C_P>0\)) indicates the existence of critical points and second-order phase transitions in the thermodynamic structure of the black hole. Comparing the curves corresponding to different values of \(\Delta\) shows that the Barrow deformation parameter plays an important regulating role in the behavior of the heat capacity. For smaller values of \(\Delta\), the thermal instability region is wider and the transition to stability occurs at larger entropies; while for larger values of \(\Delta\), the curves enter the \(C_P>0\) region more quickly and show a more stable and gentle behavior over the entire \(S_B\) interval. In other words, increasing \(\Delta\) (which indicates the strengthening of the fractal properties of the event horizon) increases the tendency of the black hole to be thermodynamically stable and reduces the intensity of thermal fluctuations.

From a physical point of view, these results indicate that the Barrow entropy-induced modifications, which are rooted in quantum gravity and the nontrivial geometry of the horizon, enrich the thermodynamic structure of the black hole and allow the emergence of diverse regimes of stability and instability. The regular dependence of \(C_P(S_B,\Delta)\) indicates the internal consistency of the model and its ability to describe phase transitions in deformed black holes, especially in the wide entropy range from near-Planck scales to macroscopic scales.

The Gibbs free energy, the enthalpy, and the internal energy in terms of the Barrow entropy are calculated as
\begin{eqnarray}
&G = \frac{1}{12 \pi ^{3/2}}\left[\frac{\pi ^2 \alpha}{b^4}  \left(2 \pi ^{3/2} \beta ^3+8 \pi  \beta ^2 S_B^{\frac{1}{\Delta +2}}+12 \sqrt{\pi } \beta  S_B^{\frac{2}{\Delta +2}}+9 S_B^{\frac{3}{\Delta +2}}\right) + \Lambda  S_B^{\frac{3}{\Delta +2}}+3 \pi  S_B^{\frac{1}{\Delta +2}}\right],\label{21-1}\\
&H = -\frac{1}{6}\beta ^3 \Lambda + \frac{1}{2} \beta + \frac{\sqrt{\pi} \alpha}{6 b^3} \left(\pi  \beta ^2 + 3 b \, S_B^{\frac{1}{\Delta +2}}\right) + \frac{1}{6 \pi^{\frac{3}{2}}} \left(- \Lambda  S_B^{\frac{3}{\Delta +2}} + 3 \pi  S_B^{\frac{1}{\Delta +2}}\right),\label{21-2}\\
&U=\frac{1}{6} \left(\frac{\pi ^{3/2} \alpha  \beta ^2}{b^3} + 3 \beta - \frac{3 \pi  \alpha \beta}{b^2} + \frac{3 \sqrt{\pi } \alpha }{b}+\frac{3 S_B^{\frac{1}{\Delta +2}}}{\sqrt{\pi }}\right).\label{21-3}
\end{eqnarray}
where it is shown that these relations depend on the Barrow entropy and the corresponding free parameters.

We now continue the discussion by relating it to the topic of the JT expansion. Therefore, the JT expansion with respect to Barrow entropy will be calculated using the following expression
\begin{equation}\label{mubar1}
 \mu = \frac{4 \left(b -\beta  \sqrt{\pi}\right) \left(2 \pi  \,b^{5}-\left(\Lambda  \,b^{5}+\alpha  \,\pi^{2} \left(3 b -2 \beta  \sqrt{\pi}\right)\right) \left(b -\beta  \sqrt{\pi}\right)^{2}\right)}{3 \sqrt{\pi}\, b \left(\pi  \,b^{4}-\left(\Lambda  \,b^{4}+\pi^{2} \alpha \right) \left(-b +\beta  \sqrt{\pi}\right)^{2}\right)}
\end{equation}
where the dependence of the JT expansion on the Barrow entropy can be observed.

\begin{figure}[h]
\begin{center}
\includegraphics[scale=0.8]{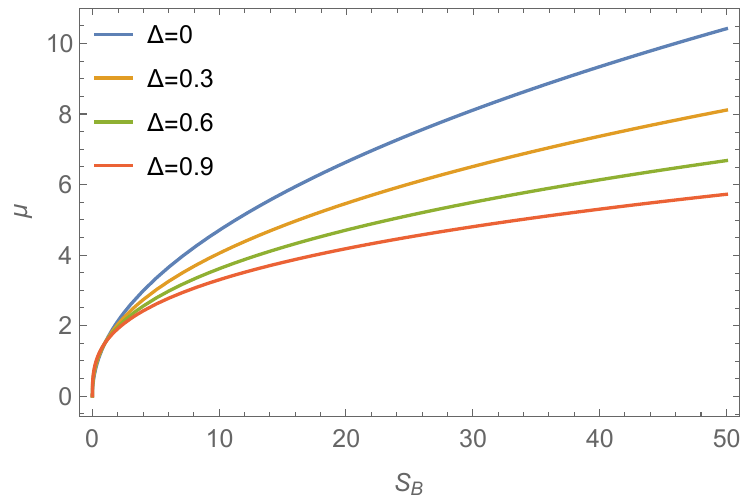}
\caption{Analyzing the JT expansion coefficient of deformed AdS-Schwarzschild black holes within the framework of Barrow entropy with respect to free parameters $\alpha=0.45, \beta=0.7$}
\label{MuB}
\end{center}
\end{figure}

In the Fig. \ref{MuB}, the JT expansion coefficient \(\mu\) for a deformed AdS-Schwarzschild black hole in the Barrow entropy framework is investigated in terms of the Barrow entropy \(S_B\) and for different values of the deformation parameter \(\Delta\). The horizontal axis \(S_B\) represents the Barrow entropy and the vertical axis \(\mu\) is the JT coefficient, which shows the rate of change of temperature with respect to pressure at constant enthalpy. Each curve corresponds to a certain value of \(\Delta\) and its changes indicate the effect of the fractal geometry of the event horizon on the thermodynamic behavior of the black hole in the JT expansion process. The behavior of the curves shows that the JT coefficient is positive for all values of \(\Delta\) and increases uniformly with increasing entropy \(S_B\). This trend indicates that the black hole is in the cooling regime (\(\mu>0\)) within the Barrow entropy framework, that is, decreasing pressure at constant enthalpy leads to a decrease in temperature. As the deformation parameter \(\Delta\) increases, the value of \(\mu\) decreases for a given entropy, such that the curve corresponding to \(\Delta=0\) (the classical case without fractal modifications) has the highest value and the curve \(\Delta=0.9\) shows the lowest value. This behavior indicates that increasing the geometric deformation of the event horizon reduces the efficiency of the JT cooling process. From a physical point of view, this decrease in the JT coefficient with increasing \(\Delta\) can be attributed to the effect of the fractal geometry of the horizon, which leads to a change in the temperature and pressure dependence in the constant enthalpy case. In fact, the Barrow entropy corrections make the black hole less thermally sensitive to pressure changes, which indicates a greater stability of the system in regimes with high deformation. In other words, the more complex and fractal the horizon structure becomes, the slower the JT process occurs and with a lower cooling rate.

In general, this figure shows that: a) The JT coefficient in the Barrow entropy framework is always positive and the black hole is in the cooling regime. b) Increasing the deformation parameter \(\Delta\) causes a decrease in the value of \(\mu\) and, consequently, a decrease in the cooling rate. c) This regular dependence \(\mu(S_B,\Delta)\) indicates the internal consistency of the thermodynamic model and the role of fractal geometry in controlling the thermal behavior of the black hole.

\section{Equation of state and thermodynamic stability}\label{III}

In this section, we examine the equations of state of a deformed AdS-Schwarzschild black hole in the framework of the Tsallis and Barrow entropies, and then analyze the thermodynamic stability conditions for each of these two frameworks. In extended phase space thermodynamics, by substituting the Eqs. \eqref{V1} and \eqref{11} into Eqs. \eqref{8} and \eqref{TBar1}, the equation of state of the AdS–Schwarzschild black hole is obtained as follows:
\begin{equation}\label{EoSTBs1}
T = \pi^{-\frac{1}{3}} P \,(6 V)^{\frac{1}{3}}
+ \frac{1}{2} \pi^{-\frac{2}{3}} (6 V)^{-\frac{1}{3}}
- \frac{2 \alpha \,(6 V)^{\frac{1}{3}}} {\left((6 V)^{\frac{1}{3}}+2 \beta \,\pi^{\frac{1}{3}}\right)^{4}}.
\end{equation}
where \(P\), \(V\), and \(T\) are pressure, volume, and temperature, respectively.

The important point is that the above equation of state has exactly the same form for both the Tsallis and Barrow entropies, and the difference between the two frameworks appears not in the form of the equation of state, but in the way the thermodynamic volume is defined. The non-extensive entropy parameters, namely \(\delta\) in the Tsallis entropy and \(\Delta\) in the Barrow entropy, are implicitly in the definition of the volume and therefore affect the behavior of the equation of state. For this reason, if we want to write the equation of state for each entropy uniquely, we must rewrite the volume \(V\) in terms of the corresponding entropy to determine the dependence of the equation of state on the specific thermodynamic structure of each entropy.

Let us now examine the thermodynamic stability of the system. Since one of the main goals of thermodynamic analysis is to determine the regions of stability and instability of a black hole, it is necessary to consider its local stability against three types of thermodynamic behavior. These behaviors include temperature fluctuations, which are related to thermal stability, pressure fluctuations, which determine mechanical stability, and adiabatic processes with constant enthalpy, which characterize JT behavior. Each of these three aspects provides a different picture of the system's response to thermodynamic perturbations and, together, completes the structure of black hole stability.

In the following, a reasonable and systematic review of each of these three levels of stability will be presented in turn.

\subsection{The thermal stability}

Thermal stability in a set at constant pressure is defined by the heat capacity at constant pressure as Eq. \eqref{10}. The sign of this quantity directly determines the state of the system's response to temperature fluctuations. If $C_P>0,$ the system is locally stable to thermal fluctuations, while $C_P<0$ it indicates thermal instability. Furthermore, divergence of \(C_P\) usually indicates the occurrence of a critical point or a second-order phase transition.

In the Tsallis entropy framework, as shown in the Fig. \ref{CpT}, the heat capacity \(C_P\) becomes smoother and more stable with increasing deformation parameter \(\delta\) and the divergences are shifted towards larger values of entropy \(S_T\). This trend shows that increasing \(\delta\) causes the expansion of the thermodynamic stability region and the reduction of the intensity of thermal instabilities.

In the Barrow entropy framework, the behavior of heat capacity \(C_P\) in terms of entropy \(S_B\) and deformation parameter \(\Delta\) also shows a similar pattern. For smaller values of \(\Delta\), sharp divergences are observed in the region of lower entropies, while with increasing \(\Delta\), the curves become smoother and the divergences shift towards larger values of \(S_B\). These changes indicate that the fractal geometry of the event horizon, which is enhanced by increasing \(\Delta\), moderates the thermal fluctuations and increases the thermodynamic stability of the black hole, as shown in Fig. \ref{CpB}.

We note that in the previous corresponding sections a more detailed examination of the heat capacity in terms of the Tsallis and Barow entropies has been carried out with respect to the non-extensive deformation parameters $\delta$ and $\Delta$ in Figs. \ref{CpT} and \ref{CpB}.

\subsection{The mechanical stability}
Here we address mechanical stability, which must be examined through the behavior of the aforementioned equation of state. This type of stability is related to isothermal compressibility, which is defined as
\begin{equation}\label{kappa1}
\kappa_T = -\frac{1}{V}\left(\frac{\partial V}{\partial P}\right)_T,
\end{equation}
where the necessary condition for mechanical stability is that $\kappa_T>0$, which can be equivalently written as
\begin{equation}\label{kappa2}
\left(\frac{\partial P}{\partial V}\right)_T < 0,
\end{equation}
where means that, the region of parameter space where the isothermal derivative of pressure with respect to volume becomes positive, $\left(\frac{\partial P}{\partial V}\right)_T > 0$, will be mechanically unstable and can correspond to spinodal regions in the phase structure of the system. As a result, accurate extraction and analysis of the equation of state plays a fundamental role in identifying phase transitions, critical points, and stability boundaries.

Now, using the obtained equation of state \eqref{EoSTBs1}, we write the pressure in terms of the non-expansive entropies of Tsallis and Barrow, and after algebraic calculations, The derivative of pressure with respect to the corresponding entropy divided by the derivative of volume with respect to the same entropy at constant temperature is obtained for Tsallis and Barrow, respectively, as follows:
\begin{eqnarray} \label{p2vt1}
&\frac{\partial_{S_T} P}{\partial_{S_T} V} \Big|_{T} = \frac{\sqrt{\pi}}{16} \left(\frac{S_T}{\gamma }\right)^{-\frac{3}{2 \delta }} \left[-\frac{2 \pi  \alpha \left(\frac{S_T}{\gamma }\right)^{\frac{1}{2 \delta }}}{\left(\sqrt{\pi } \beta +\left(\frac{S_T}{\gamma }\right)^{\frac{1}{2\delta }}\right)^5} - 2 \sqrt{\pi }\, T \left(\frac{S_T}{\gamma }\right)^{-\frac{1}{2 \delta}} + \left(\frac{S_T}{\gamma }\right)^{-\frac{1}{\delta }}\right],
\end{eqnarray} 
and
\begin{eqnarray}  \label{p2vbar1}
&\frac{\partial_{S_B} P}{\partial_{S_B} V} \Big|_{T}= -\frac{\pi^{{3}/{2}} \alpha}{8 \,S_B^{\frac{2}{2+\Delta}} \left(\beta  \sqrt{\pi} + S_B^{\frac{1}{2+\Delta}}\right)^{5}} + \frac{\sqrt{\pi}}{16 \,S_B^{\frac{5}{2+\Delta}}} - \frac{\pi \, T}{8\, S_B^{\frac{4}{2+\Delta}}},
\end{eqnarray} 
where, the stability condition \eqref{kappa2} becomes a lower bound for the temperature at both entropies. After solving the inequality with respect to $T$, the thermodynamic stable region is obtained as follows for the Tsallis and Barrow entropies, respectively:
\begin{eqnarray}
\begin{aligned}
\left\{ \begin{array}{cc}
                T > T_{c_T} = \frac{5 \sqrt{\pi}\, \beta \left(\frac{S_T}{\gamma}\right)^{\frac{3}{2 \delta}} +10 \pi^{{3}/{2}} \beta^{3} \left(\frac{S_T}{\gamma}\right)^{\frac{1}{2 \delta}} + \pi^{{5}/{2}} \beta^{5} \left(\frac{S_T}{\gamma}\right)^{-\frac{1}{2 \delta}} + 10 \pi  \left(\beta^{2}-\frac{\alpha}{5}\right) \left(\frac{S_T}{\gamma}\right)^{\frac{1}{\delta}} + \left(\frac{S_T}{\gamma}\right)^{\frac{2}{\delta}} + 5 \pi^{2} \beta^{4}}{2 \sqrt{\pi}\, \left(\beta  \sqrt{\pi} + \left(\frac{S_T}{\gamma}\right)^{\frac{1}{2 \delta}}\right)^{5}}, \\
                S_T > 0,  \\
                \end{array} \right.
\end{aligned}
\end{eqnarray}
and
\begin{eqnarray}
\begin{aligned}
\left\{ \begin{array}{cc}
              T > T_{c_B} = \frac{\pi^{{5}/{2}} \beta^{5} \, S_B^{-\frac{1}{2+\Delta}} + 10 \pi^{{3}/{2}} \beta^{3} \, S_B^{\frac{1}{2+\Delta}} + 2 \pi \left(5 \beta^{2}-\alpha\right) \, S_B^{\frac{2}{2+\Delta}} + 5 \sqrt{\pi}\,\beta S_B^{\frac{3}{2+\Delta}} + S_B^{\frac{4}{2+\Delta}} + 5 \pi^{2} \beta^{4}}{2 \sqrt{\pi}\, \left(\beta  \sqrt{\pi}+S_B^{\frac{1}{2+\Delta}}\right)^{5}}, \\
              S_B > 0, 
                \end{array} \right.
\end{aligned}
\end{eqnarray}
where $T_{c_T}$ and $T_{c_B}$ are introduced as the critical stability temperatures of the Tsallis and Barrow entropies, respectively, which are a function of the entropies and the corresponding free parameters. Thus, functions $T_{c_T}$ and $T_{c_B}$ determine the boundary between the stable and unstable thermodynamic regions. A comparison of the above two relations shows that the modifications due to the generalized entropies significantly change the thermodynamic stability structure of the black hole. In the Tsallis formalism, the stability boundary is controlled by powers of $\left(\frac{S_T}{\gamma }\right)^{\frac{1}{\delta}}$, which reflects the non-expansive nature of the entropy. In contrast, in the Barrow formalism, fractional powers of $S_B^{\frac{1}{\Delta +2}}$ appear, which originate from the fractal geometry and deformation of the black hole horizon. Hence, these two approaches are expected to predict different stability regions in the parameter space and consequently produce distinct phase behavior for AdS-Schwarzschild black holes. We note that $S_T > 0$ and $S_B > 0$ are the physical conditions for the positiveness of the Tsallis and Barrow entropies, which determine the domain of validity of AdS-Schwarzschild black hole thermodynamics. In the region $S > 0$ and $T > T_c$ (for both entropies), the black hole is mechanically stable and exhibits a good thermodynamic response to small fluctuations, while for $S < 0$ and $T < T_c$ (for both entropies), the system enters an unstable region and the mechanical stability condition is violated. Therefore, the obtained results show that the corrections due to the Tsallis and Barrow entropies not only change the thermodynamic quantities, but also shift the stability boundaries of the system and can play an important role in the phase structure and phase transitions of AdS-Schwarzschild black holes.

\subsection{Stability of JT Behavior}

The JT analysis in black hole thermodynamics is complementary to the two standard stability analyses, namely thermal stability based on the heat capacity \(C_P\) and mechanical stability based on isothermal compression \(\kappa_T\), which examines the behavior of a system in adiabatic processes with constant enthalpy and is defined by the JT coefficient of Eq. \eqref{12}. The sign of this coefficient determines whether the black hole is cooling or heating in adiabatic paths, so that if \(\mu>0\), the decrease in pressure causes a decrease in temperature and the system is in the cooling regime, on the contrary, if \(\mu<0\), the decrease in pressure causes an increase in temperature and the system is in the heating regime.

In the analyses performed for the two entropies of Tsallis and Barrow, the behavior of the JT coefficient provides a complementary picture of the thermodynamic stability structure. In both frameworks, the JT coefficient is positive over a wide range of entropy, indicating that the black hole is generally in the cooling regime. However, the cooling intensity is directly affected by non-extensive parameters, so that in the Tsallis entropy, increasing the deformation parameter \(\delta\) reduces the cooling rate, and in the Barrow entropy, increasing the geometric parameter \(\Delta\) has the same effect. This behavior suggests that statistical (in Tsallis) and geometric-fractal (in Barrow) modifications modulate the thermal sensitivity of the black hole to pressure changes.

Overall, the JT analysis shows that although thermal and mechanical stability determine the main structure of the thermodynamic behavior, the cooling/heating behavior in adiabatic paths also plays an important role in fully understanding the stability of a black hole. This analysis becomes especially important in non-expansive models, since deformation parameters can expand or restrict the cooling regions, thereby affecting the phase structure and thermal response of the system.

\section{Conclusion}\label{IV}

In this paper, the thermodynamics of a deformed AdS–Schwarzschild black hole was comprehensively investigated within the framework of two generalized entropy approaches of Tsallis and Barrow. First, the main thermodynamic quantities including mass, temperature, heat capacity, Gibbs free energy, enthalpy and internal energy were extracted and then the behavior of this system in the extended phase space was analyzed by studying the JT expansion process, the equation of state and various thermodynamic stability conditions. This framework allowed for a direct comparison of the effects of two types of generalized entropy on the thermodynamic behavior of a deformed black hole and provided a comprehensive picture of the role of geometric and statistical corrections in the thermodynamics of black holes.

The obtained results showed that the geometric deformation parameter plays a decisive role in the thermodynamic structure of a black hole. The heat capacity analysis showed that the divergences \(C_P\) correspond to the occurrence of second-order phase transitions and that the location of these critical points is directly dependent on the deformation parameters. In both the Tsallis and Barrow frameworks, increasing the deformation parameter causes the critical points to shift, the unstable regions to decrease, and the thermal stability region to expand, although the manner in which these changes occur is different in the two approaches. This behavior suggests that the modifications resulting from the generalization of the entropy relation can significantly change the structure of phase transitions and the thermal response of the black hole.

The study of the Gibbs free energy, enthalpy, and internal energy also showed that the resulting thermodynamic relations in both frameworks are in good internal consistency and provide a richer picture than the standard BH thermodynamics. These results indicate that generalized entropies are able to take into account some of the effects of gravitational corrections and the microscopic structure of the event horizon in the thermodynamic description of black holes, and thus provide a more suitable framework for studying gravitational systems with long-range interactions.

Next, the JT expansion process was studied as a tool to investigate the cooling and heating behavior of a black hole in the extended phase space. The results showed that the JT coefficient remains positive in both frameworks, and therefore the black hole is in the cooling region, but the intensity of this process depends on the deformation parameters. In the Tsallis entropy framework, a decrease in the deformation parameter causes a significant increase in the JT coefficient and enhances the cooling process, while in the Barrow entropy framework, an increase in the fractal parameter causes a gradual decrease in the JT coefficient and, consequently, a decrease in the cooling intensity. This difference shows that statistical and geometric modifications each follow a different mechanism in controlling the thermodynamic response of the black hole.

Also, by deriving the equation of state and simultaneously examining thermal stability, mechanical stability, and stability of the JT process, the stability structure of the black hole was fully analyzed. It was found that the condition of positive heat capacity, positive isothermal compressibility, and the sign of the JT coefficient are three complementary criteria for determining the stable and unstable regions of the system. In addition, the mechanical stability analysis showed that the modifications resulting from generalized entropies shift the stability boundaries and introduce new critical temperatures. Therefore, geometric deformation and generalization of the entropy relation not only modify the values of thermodynamic quantities, but also significantly change the phase space structure, stability boundaries, and phase transition characteristics of the black hole.

One of the most important achievements of this research is the direct comparison of the two generalized entropy frameworks of Tsallis and Barrow in describing the thermodynamics of a deformed black hole. Although both approaches provide a more complete description than standard thermodynamics, their physical origins are different. Tsallis entropy is based on a statistical generalization of non-extensive systems and the role of long-range interactions, while the Barrow entropy takes into account the modifications due to the fractal structure of the event horizon and the quantum effects of the geometry of space-time. This fundamental difference in physical origins also manifests itself in the behavior of the heat capacity, the equation of state, the stability boundaries, and the JT process. In particular, the Tsallis framework is more sensitive to changes in the deformation parameter and produces a stronger thermodynamic response in the cooling process, while the Barrow framework provides a smoother, more stable, and more controlled behavior, indicating the role of the horizon's fractal geometry in modulating the thermodynamic responses. These results indicate that each of these two approaches describes a different aspect of black hole physics, and their simultaneous use can provide a more comprehensive picture of the effects of modified gravity and quantum corrections on black hole thermodynamics.

In general, the results of this research show that space-time geometric deformation and generalization of the entropy relation are two independent but complementary mechanisms in modifying the thermodynamics of black holes, such that the former changes the effective geometry of space-time, while the latter modifies the statistical and microscopic description of the event horizon. The interaction of these two mechanisms reveals a richer structure of phase transitions, stability regions, equation of state and thermodynamic processes than standard thermodynamics. From this point of view, the framework presented in this paper is not only a suitable tool to study the thermodynamics of deformed black holes, but it can also be a basis for investigating the effects of modified gravity, quantum corrections and the microscopic structure of the event horizon in other gravitational systems.

Finally, this research provides an integrated framework for simultaneously investigating the effects of geometric deformation, generalized entropies, equation of state, thermodynamic stability, and the JT process in AdS–Schwarzschild black holes. This framework can be a suitable basis for future research in the field of studying JT inversion curves, critical behavior and critical points, thermodynamic geometry, generalization to charged and rotating black holes, as well as investigating these phenomena in other modified gravity theories and dark energy models.


\bibliographystyle{apsrev4-2}
\bibliography{DeformRefs}

\end{document}